\documentclass[12pt]{article}
\usepackage{amsfonts}
\usepackage{amsmath}
\usepackage{amssymb}
\usepackage{graphicx}
\usepackage{epsfig}

\begin{document}

\begin{center}
{ \large \bf From Topological to Topologically Massive Gravity through the Gauge Principle}
\end{center}


\vspace{12pt}

\begin{center}
{\large {\em Giandomenico Palumbo}}
\end{center}

\vspace{6pt}

\begin{center}
Center for Nonlinear Phenomena and Complex Systems,\\
Universit´e Libre de Bruxelles,\\ CP 231, Campus Plaine, B-1050 Brussels, Belgium\\
E-mail: giandomenico.palumbo@ulb.ac.be
\end{center}

\begin{center}
	13th May, 2019
\end{center}

\vspace{6pt}

\begin{center}
This essay has received an Honorable Mention from the Gravity Research Foundation - 2019 Awards for Essays on Gravitation.
\end{center}

\vspace{6pt}

\vspace{6pt}

\begin{abstract}
It is well known that three-dimensional Einstein's gravity without matter is topological, i.e. it does not have local propagating degrees of freedom.
The main result of this work is to show that dynamics in the gravitational sector can be induced by employing the gauge principle on the matter sector. This is described by a non-dynamical fermion model that supports a global gauge symmetry. 
By gauging this symmetry, a vector-spinor field is added to the original action to preserve the local gauge invariance.
By integrating out this spin-3/2 field, we obtain a gravitational Chern-Simons term that gives rise to local propagating degrees of freedom in the gravitational sector. This is defined, after the gauging, by topologically massive gravity.
\end{abstract}

\vspace{4 cm}

\vspace{6pt}

\vfill
\newpage

\textbf{Introduction}. 
The quantization of gravity is one of the most profound and challenging open problems of modern physics \cite{Kiefer}.
One possible way to formalize the problem is to describe gravity in terms of a classical gauge theory and then to try to quantize the theory by following the procedure developed for the Yang-Mills theories \cite{Blagojevic}.
These gauge theories are based on the gauge principle, which allows us to introduce gauge potentials in the fermionic sector of Standard Model by gauging the global gauge symmetries of the fermion fields.

The gauge-theory approach for gravity started in a seminal paper by Utiyama \cite{Utiyama}, and it has been developed and extended in further relevant works by even including a cosmological constant \cite{Blagojevic,MM}.
However, differently from the Yang-Mills theories where the gauge potentials are related to internal gauge symmetries (flavour space) and the flat background is fixed, in the classical gauge theory of gravity the fundamental degrees of freedom are the spin connection and tetrads, which are related to the dynamics of spacetime. Thus, the quantization of gravity clearly requires a sort of quantization of spacetime, which represents a challenging task to achieve.
Luckily, in three-dimensional spacetime, Einstein's gravity without matter is topological \cite{Townsend}, namely it does not have local propagating degrees of freedom and implies, for instance, the absence of gravitational waves. The corresponding gauge theory is metric-independent and coincides with a suitable Chern-Simons theory \cite{Witten}. This considerably simplifies the problem of the quantization of gravity even if the presence of local propagating degrees of freedom would allow us to figure out what kind of spacetime quantization we should implement in higher dimensions.
For this reason, there have been proposed several extensions of standard Einstein-Hilbert (EH) action, in which gravitons can propagate. Among many models, topologically massive gravity is one of the most famous and studied in literature \cite{Jackiw}.
Here, a gravitational Chern-Simons term is added to the EH action. This gives rise to a massive and propagating graviton. Besides the high-energy physics context, similar lower-dimensional gravitational models have also found interesting applications in condensed-matter physics \cite{Hughes,Palumbo1}.

In this work we will show how the gauge principle applied to fermion matter coupled to curved spacetime can induce local propagating degrees of freedom in topological gravity. This provides a novel connection between gauge theory and the emergence of dynamics in gravity and matter \cite{Palumbo2,Palumbo3}.

\vspace{0.6cm}

\textbf{Topological model}.
We start defining the gravitational sector of our model, given by the three-dimensional Einstein-Hilbert action $S_{EH}$. We employ the first-order formalism, where the spacetime variables are the dreibein $e_{\mu}^{a}$ and (dual) spin connection $\omega_{\mu}^{a}$ \cite{Blagojevic}. In this way, the gravitational action is given by
\begin{equation}\label{EH}
  S_{EH}=-\frac{1}{8\pi G}\int d^{3}x\, \epsilon^{\mu\nu\lambda} e^{a}_{\mu} \left[\partial_{\nu}\omega_{\lambda a}-\partial_{\lambda}\omega_{\nu a}+\frac{1}{2}\,\epsilon_{abc}\left(\omega^{b}_{\nu}\omega^{c}_{\lambda}-\omega^{b}_{\lambda} \omega^{c}_{\nu}\right) \right],
\end{equation}
where  $a,b,c=\{0,1,2\}$ and $\mu,\nu,\lambda=\{0,1,2\}$, $\epsilon^{\mu\nu\lambda}$ is the Levi-Civita symbol and $G$ is the Newton's constant. This theory does not have local propagating degrees of freedom because is topological and can be written as a Chern-Simons theory as shown in Refs \cite{Townsend, Witten}. Nevertheless, in presence of a negative cosmological constant $\Lambda$ the Einstein's theory supports black-hole solutions \cite{Zanelli}. We have fixed to zero $\Lambda$ in order to have a well-defined flat spacetime limit for the matter field that we are going to introduce.

The matter sector is defined by the following action term
\begin{equation}\label{matter}
S_{M}=l\int d^{3}x\, \epsilon^{\mu\nu\lambda} D_{\mu}\bar{\chi}\,\hat{\gamma}_{\nu}\,D_{\lambda}\chi,
\end{equation}
where $l$ is a dimensionful constant parameter, $\chi$ is spinor field describing a neutral fermion and $\bar{\chi}=\chi^{\dagger} \gamma_{0}$ its conjugate, $D_{\mu}=\partial_{\mu}+\omega_{\mu}$ is the covariant derivative, with $\omega_{\mu}=(1/2)\omega^{a}_{\mu}\gamma_{a}$ and $\hat{\gamma}_{\nu}=e_{\mu}^{a}\gamma_{a}$, where $\gamma_{a}$ are the Dirac matrices.

This action is completely metric independent and differently from the Dirac theory, here fermions are not dynamical similarly to the four - dimensional model proposed in Ref. \cite{Palumbo3}.
This can be easily seen by calculating the corresponding equations of motion, given by
\begin{eqnarray}
  \epsilon^{\mu\nu\lambda} D_{\mu}(\hat{\gamma}_{\nu}\,D_{\lambda}\chi)=0, \hspace{0.4cm} \epsilon^{\mu\nu\lambda} D_{\lambda}(D_{\mu}\bar{\chi}\,\hat{\gamma}_{\nu})=0,
\end{eqnarray}
where the left sides of both equations are identically null in the flat spacetime limit: $e^{a}_{\mu}\rightarrow \delta_{\mu}$, $\omega^{a}_{\mu}\rightarrow 0$.
Thus, our total action, given by
\begin{eqnarray}\label{topological}
S=S_{EH}+S_{M},
\end{eqnarray}
is dynamically trivial.

Importantly, the neutral spinor field has a global gauge symmetry, given by
\begin{equation}
\chi\rightarrow \chi + \xi, \hspace{0.4cm} \bar{\chi}\rightarrow \bar{\chi} + \bar{\xi},
\end{equation}
with $\xi$ and $\bar{\xi}$ constant spinors. This fermionic shift symmetry clearly leaves $S_{M}$ invariant
(see Ref. \cite{Bellazzini} for a different application of the fermionic shift symmetry in high-energy physics).

\textbf{Gauging the fermionic shift symmetry}.
We are now ready to show how dynamical degrees of freedom can emerge from the topological theory (\ref{topological}).
We employ the gauge principle as a guide line to introduce a novel field in the system.
In fact, by gauging the global shift symmetry, we can preserve the local gauge invariance by introducing a novel vector-spinor field $\psi_{\mu}$ that satisfies the following gauge transformations in curved spacetime \cite{Love}
\begin{eqnarray}
\psi_{\mu}\rightarrow \psi_{\mu}+D_{\mu}\xi, \hspace{0.4cm} \bar{\psi}_{\mu}\rightarrow \bar{\psi}_{\mu}+D_{\mu}\bar{\xi},
\end{eqnarray}
with $\bar{\psi}_{\mu}=\psi_{\mu}\gamma_{0}$. In this way, with the following replacements
\begin{eqnarray}
  D_{\mu}\chi\rightarrow D_{\mu}\chi-k\psi_{\mu}, \hspace{0.4cm} D_{\mu}\bar{\chi}\rightarrow D_{\mu}\bar{\chi}-k\bar{\psi}_{\mu},
\end{eqnarray}
with $k$ a dimensionful parameter, the matter sector (\ref{matter}) becomes gauge invariant with respect to the local fermionic shift symmetry.
The covariant action for $\psi_{\mu}$ is given by
\begin{eqnarray}\label{RS}
  S_{RS}=\int d^{3}x\, \epsilon^{\mu\nu\lambda}\bar{\psi}_{\mu}D_{\nu}\psi_{\lambda},
\end{eqnarray}
which is nothing but the three-dimensional Rarita-Schwinger action for spin-3/2 fermions \cite{Tucker}. Notice, the above action term together with $S_{EH}$ describe topological supergravity. However, we do not claim any fundamental role for supersymmetry here because $S_{RS}$ has been introduced for a different purpose.
Importantly, even this term is topological in three dimensions unless a mass term is introduced.
This mass term indeed appears in the matter sector of our gauged model as we show now.
We have that
\begin{align}
 S_{M}^{g}=\int d^{3}x\, \epsilon^{\mu\nu\lambda} \left[l(D_{\mu}\bar{\chi}-k\bar{\psi}_{\mu})\,\hat{\gamma}_{\nu}\,(D_{\lambda}\chi-k\psi_{\lambda})+\bar{\psi}_{\mu}D_{\nu}\psi_{\lambda}\right],
\end{align}
where we can recognize the following term
\begin{align}
S_{m}=m\int d^{3}x\, \epsilon^{\mu\nu\lambda} \bar{\psi}_{\mu}\,\hat{\gamma}_{\nu}\psi_{\lambda},
\end{align}
with $m=l k^{2}$, which represents the mass term of the Rarita-Schwinger theory \cite{Deser}.
This is the first result of gauging the fermionic shift symmetry in our model.
The mass term, which is parity-odd, has relevant consequences for the gravitational sector as we will show in the next section.

\textbf{Induced topologically massive gravity}.
As shown in Refs \cite{Deser,Dereli}, the presence of a mass term for the spin-3/2 field allows fermions to become dynamical.
Importantly, the equations of motion for $\psi_{\mu}$ shows that only one component of the vector-spinor propagates and satisfies the massive Dirac equation, where the Dirac mass is given by $m$ \cite{Dereli}.
This Dirac fermion is still coupled to the curved background through the spin connection $\omega_{\mu}$ and the dreibein $e_{\mu}^{a}$.
By focusing on the low-energy limit of our model, we can integrate out the Dirac field, such that an effective action, that depends only on the spin connection, is induced.
At one loop, the dominant term of the effective action is given by the gravitational Chern-Simons term
 \cite{Vuorio,Rao,Vassilevich}
 \begin{align}
 S_{CS}= \frac{1}{192 \pi}\frac{m}{|m|} \int d^{3}x\, \epsilon^{\mu\nu\lambda} {\rm tr}\, \left(\omega_{\mu}\partial_{\nu}\omega_{\lambda}+\frac{2}{3}
 \omega_{\mu}\omega_{\nu}\omega_{\lambda}\right),
 \end{align}
where the trace is taken over the gauge index $a$.
This term has profound consequences for the three-dimensional graviton.
Our gravitational sector is now described by the following action
\begin{eqnarray}
 S_{grav}=S_{EH}+S_{CS},
\end{eqnarray}
which is the action of topologically massive gravity \cite{Jackiw}.
Differently, from the purely Einstein-Hilbert action, this theory breaks parity and the graviton is massive and propagates.
This can be easily show by linearizing the topologically massive theory around the flat spacetime.

Importantly, even if the gauged model has more degrees of freedom than the ungauged one, our low-energy theory contains only the original variables, namely $\omega_{\mu}^{a}$, $e_{\mu}^{a}$ and $\chi$, where the gravitational sector contains now local propagating degrees of freedom. 

In conclusion, starting from topological gravity with topological matter, we have shown that the gauge principle allows us to introduce a new massive spin-3/2 field, a reminiscent of supergravity theories, that can be integrated out in the low energy regime.
This gives rise to a novel effective gravitational theory where now the graviton has dynamics.
Our work shows a new and non-trivial relation between gauge theory and emergent dynamics in topological gravity.

\textbf{Acknowledgments}. G. P. acknowledges ERC Starting Grant
TopoCold for financial support.

\end{document}